\documentclass[aps,prl,twocolumn, superscriptaddress,floatfix]{revtex4-1}
\usepackage{graphicx}
\usepackage{epstopdf}
\usepackage{bm, amsmath, amssymb}
\usepackage{pdfpages}
\usepackage{hyperref}
\usepackage{xcolor}

\begin{document}

\title{Tunability of microwave transitions as a signature of coherent parity mixing effects in the Majorana-Transmon qubit}

\author{Eran Ginossar}

\affiliation{Advanced Technology Institute and Department of Physics, University of Surrey, Guildford, GU2 7XH, United Kingdom}

\author{Eytan Grosfeld}

\affiliation{Department of Physics, Ben-Gurion University of the Negev, Be'er-Sheva 84105, Israel}

\begin{abstract}
Coupling Majorana fermion excitations to coherent external fields is an important stage towards their manipulation and detection. We analyse the charge and transmon regimes of a topological nano-wire embedded within a Cooper-Pair-Box, where the superconducting phase difference is coupled to the zero energy parity states that arise from Majorana quasi-particles. We show that at special gate bias points, the photon-qubit coupling can be switched off via quantum interference, and in other points it is exponentially dependent on the control parameter $E_J/E_C$. As well as a probe for topological-superconductor excitations, we propose that this type of device could be used to realise a tunable high coherence four-level system in the superconducting circuits architecture.
\end{abstract}

\date{\today}

\maketitle
\emph{Introduction.}--- Photons are used to control and measure qubits in a diverse range of qubit systems from natural atoms to semiconducting and superconducting solid-state architectures \cite{pellizzari1995decoherence,rauschenbeutel1999coherent,imamog1999quantum,jaksch2000fast,
raimond2001manipulating,wallraff2004strong,grosfeld2007predicted,houck2008controlling}. The properties of solid-state qubits and resonators and coupling to photons can be engineered to some degree by design and this enables the combination of several different subsystems to form hybrid devices that take advantage of their relative strengths \cite{kubo_strong_coupling_2010,schuster_high_2010}. In turn, this leads to new methods of probing physical systems and, where highly quantum coherent subsystems are involved, to establishing control over their quantum variables.

This approach is of particular interest in the context of topological Josephson junctions, where theoretical predictions~\cite{kitaev2007unpaired,lutchyn2010majorana,oreg2010helical} supported by experimental progress~\cite{mourik2012signatures,das2012zero} indicate the presence of the highly sought-after Majorana zero-energy modes~\cite{read2000paired,moore1991nonabelions} localized on the wire around its endpoints. To probe these excitations with microwave photons requires a hybrid device formed by bridging a Josephson junction using a suitably prepared nanowire. The minimal quantum description of the device involves an effective {\em parity fermion} degree of freedom, that arises from the Majorana quasi-particles on the wire, and represents the parity of the total charge. This fermion is coupled to the phase and charge degrees of freedom of the superconducting circuit. Anticipating experimental progress in these directions, a central question that needs to be addressed is what is the unique spectroscopic signature of the charge-neutral Majorana zero modes in such devices? Also, can one establish control over the parity fermion, demonstrating its relevance for quantum information processing?

\begin{figure}[b]
\centering
\includegraphics[scale=0.6]{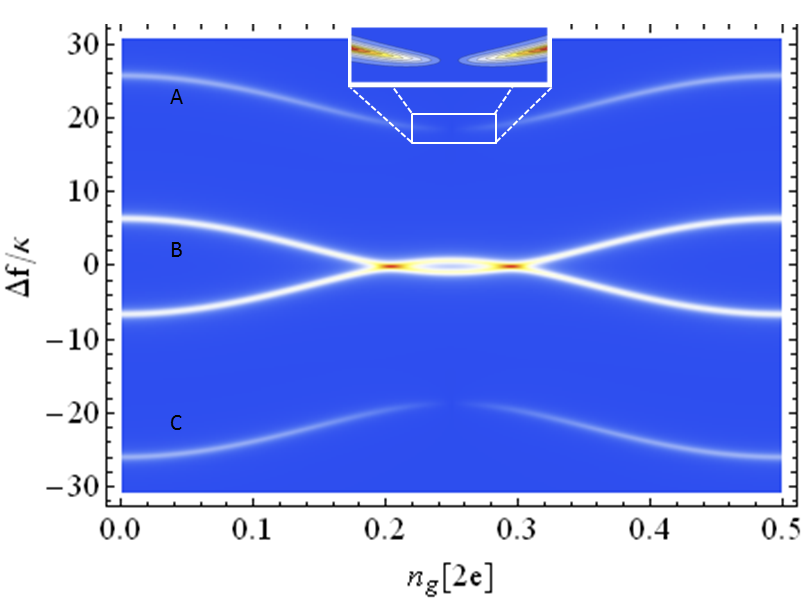}
\caption{(Color online) Spectroscopy of a Majorana-Transmon qubit device. The microwave transition frequencies ($\Delta f$) of the system as function of the offset charge $n_g$ for the different transition paths within the lower part of the spectrum (see Fig.~\ref{fig:setup}b). The upper (A) and lower (C) lines are associated with the usual transmon frequencies while the central lines (B) describe parity mixing effects and are a unique feature of the Majorana modes. At $n_g=1/4$ the transition amplitudes in each parity manifold interfere destructively and this gives rise to a `spectral hole' in transition (A,C). In contrast to the traditional transmon, all spectral lines are gate tunable, and the pattern can be shifted between 4 (left) and 2 (center) resonant frequencies. The frequencies are plotted relative to the average and in units of the line width $\kappa/2\pi=50\textrm{KHz}$, taking $E_C/2\pi=400\textrm{MHz}$, $E_J/E_C=27$ and $E_M/2\pi=0.5 \textrm{MHz}$.}
\label{fig:spectroscopy}
\end{figure}

\begin{figure}[t]
\centering
\includegraphics[scale=0.45]{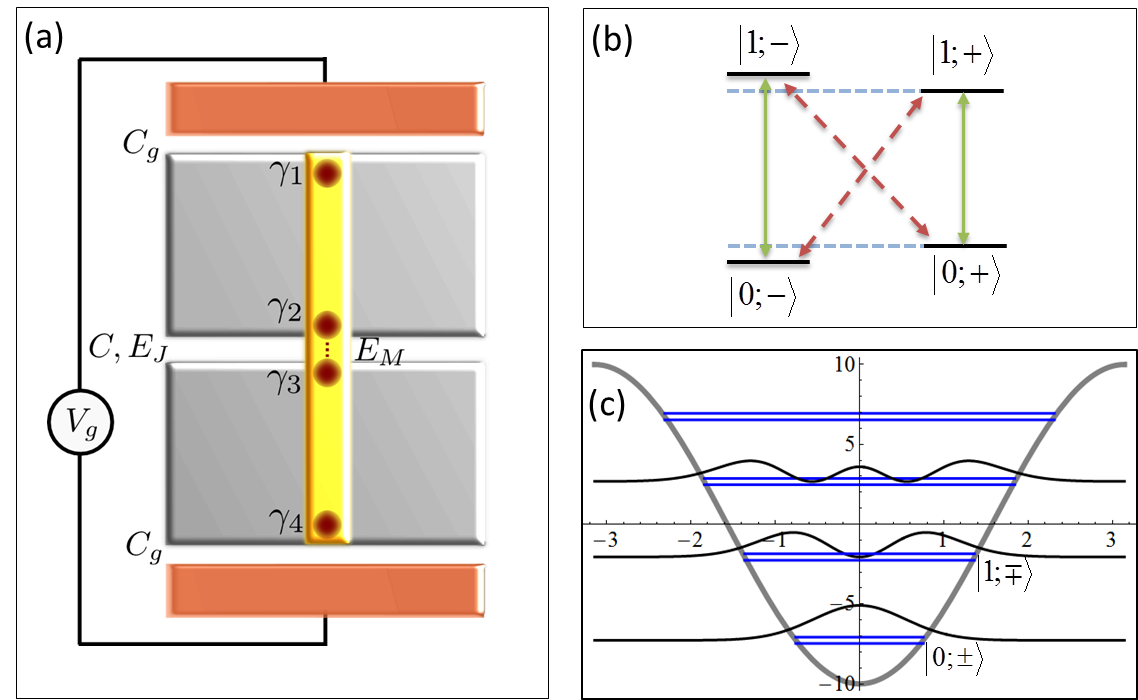}
\caption{(Color online) The experimental setup considered in the text (a). Two superconductors (grey) form a Josephson junction and are capacitively coupled to a gate (orange). A topological nanowire lying in close proximity (yellow) carries four Majorana zero modes. Two of the zero modes hybridize (dashed line) owing to single electron tunneling processes across the junction. The eigenstates, denoted $|0/1;\pm\rangle$ are coherent superpositions of the opposite fermion parity transmon states. Panel (b) shows the tunable microwave couplings scheme between the eigenstates, realising a double-$\Lambda$ system. The various optical transition strengths can be controlled by varying either $E_J$ or $n_g$. (c) Representative probability densities of the two parity doublets $(\pm)$ are plotted against the background of the dominant Josephson energy potential, for $E_C/h=0.4\,\textrm{GHz}, E_J/E_C=25, E_M/E_C=5\times 10^{-4}, n_g=0.25$. The doublet energy splittings are drawn exaggerated for visibility in panels (b,c).}
\label{fig:setup}
\end{figure}

\begin{figure}[b]
\includegraphics[scale=0.35]{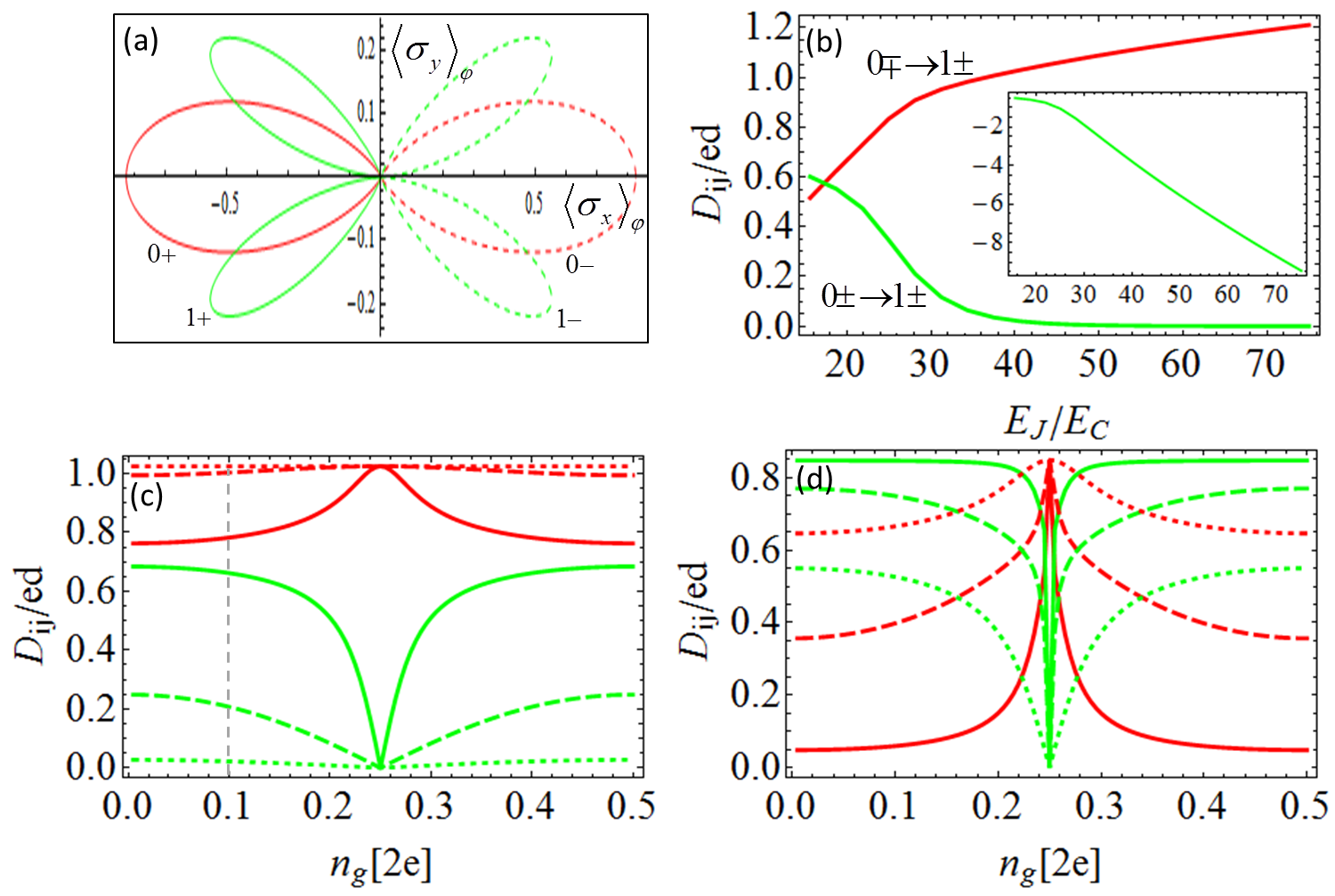}
\caption{(Color online) (a) The fermion tunneling term $H_M$ leads to a correlation between the superconducting phase difference ($\varphi$) and the parity-spinor. The $x-y$ projection of the Bloch vector representation of the parity-spinor is plotted as a function of $\varphi$ at the point of maximal hybridization $n_g=1/4$. (b) Shows the dependence of the optical transition strength $\mathcal{G}_{k,\pi\rightarrow k,\pi'}$ as function of $E_J/E_C$ with $E_C/h=0.4\,\textrm{GHz}, n_g=0.1$ and $E_M/E_C=1/400$, and the logarithm of this dependence (inset, same x-axis) shows that the suppression is exponential. (c) Shows the dependence of the dipole on the effective gate charge $n_g$ for $E_J/E_C=40$ with $E_C/h=0.4\,\textrm{GHz}$, and for $E_J/E_C=20$ (d). Three curves are drawn in each panel (c,d) to show the dependence on $E_M/E_C=2.5\times 10^{-3} (\textrm{dotted}),2.5\times 10^{-4} (\textrm{dashed}),2.5\times 10^{-5} (\textrm{solid})$. Strong parity-phase correlations arise near the avoided crossing point $n_g=1/4$ which results in a vanishing of the coupling between the $|0\rightarrow 1;\pm \rangle$ states. Deeper in the transmon regime ($E_J/E_C=40$) these correlations persist across the whole range of $n_g$ even for small values of the Majorana coupling $E_M$ which suppresses the optical coupling matrix element, see dotted line in (c).}
\label{fig:dipole}
\end{figure}

To answer these questions, in this Letter we explore the properties of a Majorana-Cooper-Pair-Box parity-charge qubit hybrid in the charge \cite{stern2006proposed,ilan2008coulomb,
ilan2009experimental,fu2010electron,van2011coulomb,golub2012charge} and transmon \cite{koch_charge-insensitive_2007} regimes. We find two key differences when compared with the traditional Cooper-Pair-Box. First, a remarkable spectroscopic pattern consisting of two to four resonant frequencies whose number and intensities are tunable via a gate offset $n_g$ (see Fig.~\ref{fig:spectroscopy}); the pattern is very different from the traditional transmon that admits exactly two resonant frequencies with non-tunable intensities. Second, an excitation spectrum consisting of multiple doublets inherited from the fermion parity states (see Fig.~\ref{fig:setup}c), whose lowest four levels can be used to form a double-$\Lambda$ system with a highly coherent almost degenerate doublet ground state. In the rest of this Letter we shall refer to this ground state doublet as the ``Majorana-Transmon qubit''. The combination of these properties allows: (i) Spectroscopic \emph{detection} of the parity fermion strengthening currently available transport signatures of Majorana fermions in a setting that avoids complications stemming from the need to attach leads; and, (ii), coherent and tunable \emph{control} over the parity fermion made possible by the highly anharmonic level structure and the tunability of transitions. As we now explain, the coupling strengths for the different transitions are determined by a quantum interference effect that arises from the presence of the coherently mixed parity fermions and reveal their presence. In particular, the parity-conserving-like transitions (green), see Fig.~\ref{fig:setup}b, can be tuned to exactly zero by the gate offset $n_g$ or exponentially suppressed by varying the ratio $E_J/E_C$ (Fig.~\ref{fig:dipole}b-d). Such tunable control allows state manipulation of the Majorana-Transmon qubit via the upper levels. These properties are highly desirable for quantum information processing in the superconducting qubit architecture which is currently based on a very weakly anharmonic level structure that limits the speed and fidelity of gate operations.

\emph{Proposed setup and description of the effective model}.--- The proposed hybrid device contains a nano-wire which is placed in proximity to the Josephson junction of a Cooper-Pair-Box (CPB) (see Fig.~\ref{fig:setup}), the latter being a prototypical charge qubit (realized as either a 2D or 3D transmon). A combination of strong spin-orbit coupling and a Zeeman gap can be used to push the wire into its topological state, provided that the chemical potential is tuned within the wire's gap. In this phase, Majorana zero modes with operators $\gamma_2$ and $\gamma_3$ are localized near the junction. Two additional distant Majorana modes are present within the superconductors, $\gamma_1$ and $\gamma_4$ (see Fig.~\ref{fig:setup}a). The operators satisfy $\gamma_i^\dagger=\gamma_i$ and $\{\gamma_i,\gamma_j\}=\delta_{ij}$. The properties of the device are determined by the interplay of the Josephson coupling $E_J$, the charging energy $E_C$, the coupling $E_M$ between the Majorana excitations and the superconducting phase difference $\varphi$ \cite{kwon2004fractional,fu2009josephson,fu2010electron,grosfeld2011observing,golub2012charge,mora2012low,dutt2013strongly}. The Hamiltonian for the hybrid junction is $H=H_0+H_M$ where
\begin{eqnarray}
	&&\nonumber H_0[n_g]=4E_C\left(\frac{1}{i}\partial_\varphi-n_g\right)^2-E_J\cos\varphi,\\
	&&H_M=i E_M \gamma_2 \gamma_3 \cos(\varphi/2). \label{eq:maj-tunneling}
\end{eqnarray}
Here $n_g$ describes the total offset charge that represents an external electrostatic gate control. The anomalous term $H_M$ is generated by coherent single-electron tunnelling processes between the two superconductors facilitated by the presence of the zero-energy Majorana modes. Its presence was predicted to lead to observable effects \cite{kwon2004fractional,nilsson2008splitting,fu2009josephson,tanaka2009manipulation,van2011coulomb,jiang2011unconventional,law2011robustness,ioselevich2011anomalous}, of which some experimental signatures were recently observed \cite{rokhinson2012fractional}. In our context, it allows coherent Rabi oscillations of unpaired electrons. We focus on the Majorana-Transmon (MT) regime, which we define as $E_M\ll E_C \ll E_J$, where the influence of charge noise is exponentially suppressed, and explore the dependence of the eigen-states and eigen-energies spectrum on $n_g$. When $E_M$ is non-zero, we find a spectrum that is composed of closely spaced doublets of transmon-like energy levels with a periodicity of $e$ compared to $2e$ for the transmon. The microwave photons couple to the charge operator and we find that even when $E_M /E_C \ll 1$  the charge matrix element $\langle i|\hat{n}|j\rangle$ that couples to the photon field is strongly dependent on $n_g$ and $E_J/E_C$ for the allowed GHz-range transitions since the bare states of the system become strongly hybridized. For completeness, we would like to mention other scenarios that were recently discussed in the literature: the $E_M=0$ and $E_M=E_C=0$ limits in \cite{hassler2011top,bonderson2011topological}; the flux qubit in \cite{pekker2013proposal}; circuit QED extensions in \cite{blais_majorana_2013,cottet_majorana_squeezing_2013}; and, photon-induced long-distance Majorana coupling in \cite{schmidt2013majorana}. In contrast, the present work offers an effective model capturing the multi-level spectrum of the charge-qubit device. Due to a particular set of useful features that we now elaborate on, this minimal scenario may also have considerable appeal as an alternative to transport based experiments.

\emph{Solution of the effective model}.--- In the topological phase, the nano-wire excitations carry a single zero energy fermion state on either side of the junction, whose occupation becomes locked to the parity of the fermion number on the same side \cite{fu2010electron,zazunov2011coulomb,golub2012charge}. By projecting on the states labelled by these fermion numbers, and focusing on a single total parity state, the Hamiltonian acquires a matrix structure
\begin{eqnarray}
	H=\left(\begin{array}{cc} H_0[n_g] & E_M \cos(\varphi/2) \\ E_M \cos(\varphi/2) & H_0[n_g] \end{array}\right), \label{eq:ham}
\end{eqnarray}
which we now diagonalize by solving the eigenvalue equation $H \chi=E\chi$ where $\chi=(f(
\varphi),g(\varphi))^T$. A crucial point is that one should take into account the requirement on the Hilbert space that $f(\varphi)$ is periodic in $\varphi$ with a periodicity $2\pi$, while $g(\varphi)$ is anti-periodic. Alternatively one can use a basis composed of solely periodic functions, but the Hamiltonian gets modified according to $H\to H'=U H U^\dagger$, with $U=\mbox{diag}\{1,e^{i\varphi/2}\}$
\begin{eqnarray}
	H'=\left(\begin{array}{cc} H_0[n_g] & \frac{E_M}{2} (1+e^{-i\varphi}) \\ \frac{E_M}{2} (1+e^{i\varphi}) & H_0[n_g+1/2] \end{array}\right). \label{eq:ham-transformed}
\end{eqnarray}
The eigen-energies of this Hamiltonian were calculated numerically employing charge eigenstates and are presented in Fig.~\ref{fig:spectrum} as function of $n_g$.

Further insight into the effect of the coupling $H_M$ on the system can be gained from diagonalising the Hamiltonian in the basis of the transmon eigenfunctions \cite{koch_charge-insensitive_2007} which will be denoted as $\Psi_k(n_g,\varphi)$ and which diagonalise $H_0 [n_g]$ and $H_0 [n_g+1/2]$, respectively. In this basis $H'$ can be shown to be approximately $2\times 2$ block-diagonal since $H_M$ mostly couples the same band pairs $\{\Psi_k(n_g,\varphi),\Psi_k(n_g+1/2,\varphi) \}$. This coupling results in an energy splitting of the order $E_M$ developing around $n_g=1/4$ (see Fig.~\ref{fig:spectrum}b). Importantly, the periodicity of the energy levels is halved, and this will affect the behavior of the system with respect to tunneling of non-equilibrium quasi-particles. Further, in the transmon regime $E_J/E_C \gg 1$ the dispersion of the transmon levels is exponentially suppressed such that the $E_M$ dominates the spectral gap and flattens it further, see Fig.~\ref{fig:spectrum}(c-d), and hence improving its famous resilience against dephasing caused by charge fluctuations.

\emph{Light-matter interaction}.--- Electromagnetic fields influence the dynamics of the MT by coupling via the dipole operator, given by $D=i d e \partial_\varphi$, where $d$ is the distance between the two superconductors. In the transformed basis $D'=UDU^\dagger$ the dipole operator acquires the form
\begin{eqnarray}
	D'=ed\left(\begin{array}{cc}
	i\partial_\varphi & \\
	& i\partial_\varphi+\frac{1}{2}\\
	\end{array}\right).
\end{eqnarray}
The parity-phase correlation has important consequences for the microwave and RF coupling between the states. The matrix elements of the dipole operator
\begin{eqnarray}\nonumber
&&	\mathcal{G}_{k,\pi\rightarrow k',\pi'}(n_g)=(ed)^{-1}\int_0^{2\pi}d\varphi \langle k,\pi|D'|k',\pi'\rangle=\\
&& \int_0^{2\pi}d\varphi\left[ f_{k,\pi}^*f_{k',\pi'}\Psi_k^*(n_g;\varphi)i\partial_\varphi\Psi_{k'}(n_g;\varphi) \right.\\
&& \left. +g_{k,\pi}^*g_{k',\pi'}\Psi_k^*(n_g+1/2;\varphi)(i\partial_\varphi+1/2)\Psi_{k'}(n_g+1/2;\varphi)\right]\nonumber
\end{eqnarray}
where $\pi,\pi'=\pm$, yields the coherent sum of two matrix elements resulting from the different fermion parities. Thus, the transition amplitude between states of same or different $\pi,\pi'$ is given by the coherent addition of the matrix elements (see Fig.~\ref{fig:setup}b). Usually, the integrals need to be evaluated numerically, but some observations can be made on general grounds. At the degeneracy point $n_g=1/4$ the parity-spinor amplitudes have equal weight of uncoupled MT components and the matrix elements for the transitions $|0;\pi\rangle\rightarrow|1;\pi'\rangle$ become
\begin{eqnarray}\label{eq:dipoleatdeg}\nonumber
&&	\mathcal{G}_{0,\pi\rightarrow 1,\pi'}(n_g=1/4)=\\ \nonumber
&&  \frac{i}{2} \int_0^{2\pi}d\varphi \left[ \Psi_0^*(n_g=1/4,\varphi)\partial_\varphi\Psi_1(n_g=1/4,\varphi)\pm \right. \\
&&  \left. \Psi_0^*(n_g=3/4,\varphi)\partial_\varphi\Psi_1(n_g=3/4,\varphi) \right]
\end{eqnarray}
where the relative sign $\pm$ depends on the choice of pair of states. At this degeneracy point, for the transition corresponding to $(-)$, a full destructive interference between the dipole matrix elements \footnote{At this point it can be shown that $\Psi_k(n_g=3/4;\varphi)=e^{i\varphi}\Psi_k^*(n_g=1/4;\varphi)$ which leads to a partial cancellation of terms in Eq.~\ref{eq:dipoleatdeg} for the transition corresponding to $(-)$ and the remaining terms are anti-symmetric.} takes place. As we move away from the special point $n_g=1/4$ the amplitudes change and the dipole coupling returns to a finite value on a scale that depends on $E_M$, see Fig.~\ref{fig:dipole}. In contrast, the `intra-band' dipole coupling $\mathcal{G}_{k,\pi\rightarrow k,\pi'}$ can be shown to be relatively small for all value of $n_g$. As $n_g$ is varied it should be possible to see the different transitions lines of Fig.~\ref{fig:setup}(b) appear and disappear as their transition rates vary. This spectroscopic pattern depends on the Majorana coupling $E_M$ and the latter may be estimated from it. The sensitivity of the dipole coupling to parameters also extends to an exponential dependence on the ratio of $E_J/E_C$ since the latter controls the relative effectiveness of the hybridization due to $H_M$. A strong on-off control of the coupling, see Fig.~\ref{fig:dipole}(b) could be achieved with a split-CPB, albeit here with a qualitatively different behavior from a recent implementation that is based on circuit QED architecture \cite{gambetta2011tunable_coupling}.

\emph{Qubit realization and control}.--- A device that realises this Hamiltonian could have substantial advantages for quantum information processing. In addition to the gate tunability of the direct microwave coupling between transmon qubit states, the two lowest MT states form a parity-energy doublet, see Fig.~\ref{fig:setup}(b), which constitutes a highly anharmonic MT qubit. The high anharmonicity is considered a benefit for qubit state manipulation as it reduces the probability of higher states being excited and shortens the duration of control pulses. The two almost degenerate ground states can be regarded as part of a $\Lambda$ type system and thus optically manipulated via transitions involving the higher doublet states. The doublet states can be described as spinors in the odd-even parity space $|k,\pm\rangle=\left(f_{k,\pm},g_{k,\pm}\right)^T$ where $k=0,1$ denotes the transmon band index and $\pm$ denotes symmetric and anti-symmetric combinations. Due to the interaction $H_M$ the effective parity and phase ($\varphi$) degrees of freedom become correlated, showing equal weight hybridization close to the crossing point $n_g=1/4$, i.e. $|f|^2-|g|^2=0$ ($\langle\sigma_z\rangle=0$). Close to this point the resulting spinor nature of the wave function can be depicted by taking the partial trace in the pseudospin space $(\langle\sigma_x\rangle,\langle\sigma_y\rangle,\langle\sigma_z\rangle)_{\varphi,\pm}=\mbox{Tr}(\rho_{k,\pm}(\varphi)\vec{\sigma})$, see Fig.~\ref{fig:dipole}(a). The $\varphi$-dependent spinor lies on the equator of the Bloch sphere due to the equal weighting. It points predominantly in the $\sigma_x$ direction since the parity of the transmon wave functions with respect to  $\varphi$ yields real matrix elements for $H_M$. The sensitivity of the states to the parameter $n_g$ leads to adiabatic control, which by tuning $n_g$ from $0$ to $1/2$ will take a system prepared for example in a state pointing predominantly in the $|k;0,0\rangle$ direction to a state close to $|k;1,1\rangle$. Microscopically, this involves splitting a Cooper-pair adiabatically between the two sides of the junction. For intermediate values the system maintains a coherent superposition of these two parity states.

\emph{Alternative scenarios}.--- Microscopic two level systems (TLS) that interact with the charge qubit generally give rise to spurious spectroscopic signatures. However, these can be differentiated from the case of Majorana fermions. In the case of a TLS that is resonantly coupled to the transmon there is only one ground state and therefore only two spectroscopic transitions to the one-excitation manifold. Correspondingly, there are only two transition frequencies that can be observed compared to up to four in the Majorana case. In contrast, if the TLS is close to degenerate in energy, the full system may possess a similar spectrum of two doublets. However, the crucial difference arises from the absence of direct coupling between the two quasi-degenerate ground states, in contrast to the case of Majorana fermions. This strongly suppresses any additional features in the spectroscopic pattern. Another possible scenario is a single electron state trapped within the oxide barrier, with a typical energy $E_{0}$, that can exchange electrons with the two superconductors. Again, the phenomenology is very different since there will always be an extra energetic cost associated with an unpaired electron within the superconductors, leading to a very high energy scale $E_{0}+\Delta$ in which such processes appear in the spectroscopic pattern (here $\Delta$ is the superconductor gap). Generally speaking, independently of the competing microscopic scenario, the Majorana fermions always satisfy: (i) the periodicity of the spectrum is coherently halved compared to the regular transmon owing to a rigidity in the relative shift $n_g\to n_g+1/2$ of the two levels that hybridize to form the doublet; and, (ii), the spectroscopic signatures do not depend on the particular realization of randomly occurring TLSs in the oxide.

To conclude, we have presented an analysis of a Majorana-Transmon qubit hamiltonian and discussed its unique spectroscopic signatures. It was shown that since the parity and transmon states are entangled, the dipole transitions become strongly dependent on the system parameters $n_g$ and $E_J/E_C$. The energy scale $E_M$ controlling the Majorana coupling can be extracted from the spectroscopic pattern as discussed above. Such spectroscopic signatures of the Majorana fermions can be differentiated from dominant impurity scenarios in the oxide barrier. We believe that these results will play an important role in the analysis of experiments which involve the embedding of topological superconductors within the superconducting qubit architecture. In addition, the set of optical selection rules that we find support the realisation of a dual-$\Lambda$ system (see Fig.~\ref{fig:setup}) with dipole allowed transitions to the higher transmon-like states of the system which can be externally controlled. This enables the preparation and manipulation of quantum states using transitions through the upper levels.

\begin{acknowledgments}

We acknowledge useful discussions with L.~DiCarlo. E.~Ginossar acknowleges support from EPSRC (EP/I026231/1). E.~Grosfeld acknowledges support from the Israel Science Foundation (Grant No.~401/12) and the European Union's Seventh Framework Programme (FP7/2007-2013) under Grant No.~303742. Both authors acknowledge support from the Royal Society International Exchanges program.

\end{acknowledgments}

\bibliographystyle{apsrev}
\bibliography{transmon_majorana}

\begin{thebibliography}{44}
\expandafter\ifx\csname natexlab\endcsname\relax\def\natexlab#1{#1}\fi
\expandafter\ifx\csname bibnamefont\endcsname\relax
  \def\bibnamefont#1{#1}\fi
\expandafter\ifx\csname bibfnamefont\endcsname\relax
  \def\bibfnamefont#1{#1}\fi
\expandafter\ifx\csname citenamefont\endcsname\relax
  \def\citenamefont#1{#1}\fi
\expandafter\ifx\csname url\endcsname\relax
  \def\url#1{\texttt{#1}}\fi
\expandafter\ifx\csname urlprefix\endcsname\relax\def\urlprefix{URL }\fi
\providecommand{\bibinfo}[2]{#2}
\providecommand{\eprint}[2][]{\url{#2}}

\bibitem[{\citenamefont{Pellizzari et~al.}(1995)\citenamefont{Pellizzari,
  Gardiner, Cirac, and Zoller}}]{pellizzari1995decoherence}
\bibinfo{author}{\bibfnamefont{T.}~\bibnamefont{Pellizzari}},
  \bibinfo{author}{\bibfnamefont{S.~A.} \bibnamefont{Gardiner}},
  \bibinfo{author}{\bibfnamefont{J.~I.} \bibnamefont{Cirac}}, \bibnamefont{and}
  \bibinfo{author}{\bibfnamefont{P.}~\bibnamefont{Zoller}},
  \bibinfo{journal}{Phys. Rev. Lett.} \textbf{\bibinfo{volume}{75}},
  \bibinfo{pages}{3788} (\bibinfo{year}{1995}).

\bibitem[{\citenamefont{Rauschenbeutel
  et~al.}(1999)\citenamefont{Rauschenbeutel, Nogues, Osnaghi, Bertet, Brune,
  Raimond, and Haroche}}]{rauschenbeutel1999coherent}
\bibinfo{author}{\bibfnamefont{A.}~\bibnamefont{Rauschenbeutel}},
  \bibinfo{author}{\bibfnamefont{G.}~\bibnamefont{Nogues}},
  \bibinfo{author}{\bibfnamefont{S.}~\bibnamefont{Osnaghi}},
  \bibinfo{author}{\bibfnamefont{P.}~\bibnamefont{Bertet}},
  \bibinfo{author}{\bibfnamefont{M.}~\bibnamefont{Brune}},
  \bibinfo{author}{\bibfnamefont{J.}~\bibnamefont{Raimond}}, \bibnamefont{and}
  \bibinfo{author}{\bibfnamefont{S.}~\bibnamefont{Haroche}},
  \bibinfo{journal}{Phys. Rev. Lett.} \textbf{\bibinfo{volume}{83}},
  \bibinfo{pages}{5166} (\bibinfo{year}{1999}).

\bibitem[{\citenamefont{Imamo\u{g}lu et~al.}(1999)\citenamefont{Imamo\u{g}lu,
  Awschalom, Burkard, DiVincenzo, Loss, Sherwin, and
  Small}}]{imamog1999quantum}
\bibinfo{author}{\bibfnamefont{A.}~\bibnamefont{Imamo\u{g}lu}},
  \bibinfo{author}{\bibfnamefont{D.~D.} \bibnamefont{Awschalom}},
  \bibinfo{author}{\bibfnamefont{G.}~\bibnamefont{Burkard}},
  \bibinfo{author}{\bibfnamefont{D.~P.} \bibnamefont{DiVincenzo}},
  \bibinfo{author}{\bibfnamefont{D.}~\bibnamefont{Loss}},
  \bibinfo{author}{\bibfnamefont{M.}~\bibnamefont{Sherwin}}, \bibnamefont{and}
  \bibinfo{author}{\bibfnamefont{A.}~\bibnamefont{Small}},
  \bibinfo{journal}{Phys. Rev. Lett.} \textbf{\bibinfo{volume}{83}},
  \bibinfo{pages}{4204} (\bibinfo{year}{1999}).

\bibitem[{\citenamefont{Jaksch et~al.}(2000)\citenamefont{Jaksch, Cirac,
  Zoller, Rolston, C{\^o}t{\'e}, and Lukin}}]{jaksch2000fast}
\bibinfo{author}{\bibfnamefont{D.}~\bibnamefont{Jaksch}},
  \bibinfo{author}{\bibfnamefont{J.~I.} \bibnamefont{Cirac}},
  \bibinfo{author}{\bibfnamefont{P.}~\bibnamefont{Zoller}},
  \bibinfo{author}{\bibfnamefont{S.~L.} \bibnamefont{Rolston}},
  \bibinfo{author}{\bibfnamefont{R.}~\bibnamefont{C{\^o}t{\'e}}},
  \bibnamefont{and} \bibinfo{author}{\bibfnamefont{M.~D.} \bibnamefont{Lukin}},
  \bibinfo{journal}{Phys. Rev. Lett.} \textbf{\bibinfo{volume}{85}},
  \bibinfo{pages}{2208} (\bibinfo{year}{2000}).

\bibitem[{\citenamefont{Raimond et~al.}(2001)\citenamefont{Raimond, Brune, and
  Haroche}}]{raimond2001manipulating}
\bibinfo{author}{\bibfnamefont{J.~M.} \bibnamefont{Raimond}},
  \bibinfo{author}{\bibfnamefont{M.}~\bibnamefont{Brune}}, \bibnamefont{and}
  \bibinfo{author}{\bibfnamefont{S.}~\bibnamefont{Haroche}},
  \bibinfo{journal}{Rev. of Mod. Phys.} \textbf{\bibinfo{volume}{73}},
  \bibinfo{pages}{565} (\bibinfo{year}{2001}).

\bibitem[{\citenamefont{Wallraff et~al.}(2004)\citenamefont{Wallraff, Schuster,
  Blais, Frunzio, Huang, Majer, Kumar, Girvin, and
  Schoelkopf}}]{wallraff2004strong}
\bibinfo{author}{\bibfnamefont{A.}~\bibnamefont{Wallraff}},
  \bibinfo{author}{\bibfnamefont{D.~I.} \bibnamefont{Schuster}},
  \bibinfo{author}{\bibfnamefont{A.}~\bibnamefont{Blais}},
  \bibinfo{author}{\bibfnamefont{L.}~\bibnamefont{Frunzio}},
  \bibinfo{author}{\bibfnamefont{R.~S.} \bibnamefont{Huang}},
  \bibinfo{author}{\bibfnamefont{J.}~\bibnamefont{Majer}},
  \bibinfo{author}{\bibfnamefont{S.}~\bibnamefont{Kumar}},
  \bibinfo{author}{\bibfnamefont{S.~M.} \bibnamefont{Girvin}},
  \bibnamefont{and} \bibinfo{author}{\bibfnamefont{R.~J.}
  \bibnamefont{Schoelkopf}}, \bibinfo{journal}{Nature}
  \textbf{\bibinfo{volume}{431}}, \bibinfo{pages}{162} (\bibinfo{year}{2004}).

\bibitem[{\citenamefont{Grosfeld et~al.}(2007)\citenamefont{Grosfeld, Cooper,
  Stern, and Ilan}}]{grosfeld2007predicted}
\bibinfo{author}{\bibfnamefont{E.}~\bibnamefont{Grosfeld}},
  \bibinfo{author}{\bibfnamefont{N.~R.} \bibnamefont{Cooper}},
  \bibinfo{author}{\bibfnamefont{A.}~\bibnamefont{Stern}}, \bibnamefont{and}
  \bibinfo{author}{\bibfnamefont{R.}~\bibnamefont{Ilan}},
  \bibinfo{journal}{Phys. Rev. B} \textbf{\bibinfo{volume}{76}},
  \bibinfo{pages}{104516} (\bibinfo{year}{2007}).

\bibitem[{\citenamefont{Houck et~al.}(2008)\citenamefont{Houck, Schreier,
  Johnson, Chow, Koch, Gambetta, Schuster, Frunzio, Devoret, Girvin
  et~al.}}]{houck2008controlling}
\bibinfo{author}{\bibfnamefont{A.~A.} \bibnamefont{Houck}},
  \bibinfo{author}{\bibfnamefont{J.~A.} \bibnamefont{Schreier}},
  \bibinfo{author}{\bibfnamefont{B.~R.} \bibnamefont{Johnson}},
  \bibinfo{author}{\bibfnamefont{J.~M.} \bibnamefont{Chow}},
  \bibinfo{author}{\bibfnamefont{J.}~\bibnamefont{Koch}},
  \bibinfo{author}{\bibfnamefont{J.}~\bibnamefont{Gambetta}},
  \bibinfo{author}{\bibfnamefont{D.~I.} \bibnamefont{Schuster}},
  \bibinfo{author}{\bibfnamefont{L.}~\bibnamefont{Frunzio}},
  \bibinfo{author}{\bibfnamefont{M.~H.} \bibnamefont{Devoret}},
  \bibinfo{author}{\bibfnamefont{S.~M.} \bibnamefont{Girvin}},
  \bibnamefont{et~al.}, \bibinfo{journal}{Phys. Rev. Lett.}
  \textbf{\bibinfo{volume}{101}}, \bibinfo{pages}{080502}
  (\bibinfo{year}{2008}).

\bibitem[{\citenamefont{Kubo et~al.}(2010)\citenamefont{Kubo, Ong, Bertet,
  Vion, Jacques, Zheng, Dréau, Roch, Auffeves, Jelezko
  et~al.}}]{kubo_strong_coupling_2010}
\bibinfo{author}{\bibfnamefont{Y.}~\bibnamefont{Kubo}},
  \bibinfo{author}{\bibfnamefont{F.~R.} \bibnamefont{Ong}},
  \bibinfo{author}{\bibfnamefont{P.}~\bibnamefont{Bertet}},
  \bibinfo{author}{\bibfnamefont{D.}~\bibnamefont{Vion}},
  \bibinfo{author}{\bibfnamefont{V.}~\bibnamefont{Jacques}},
  \bibinfo{author}{\bibfnamefont{D.}~\bibnamefont{Zheng}},
  \bibinfo{author}{\bibfnamefont{A.}~\bibnamefont{Dréau}},
  \bibinfo{author}{\bibfnamefont{J.-F.} \bibnamefont{Roch}},
  \bibinfo{author}{\bibfnamefont{A.}~\bibnamefont{Auffeves}},
  \bibinfo{author}{\bibfnamefont{F.}~\bibnamefont{Jelezko}},
  \bibnamefont{et~al.}, \bibinfo{journal}{Phys. Rev. Lett.}
  \textbf{\bibinfo{volume}{105}}, \bibinfo{pages}{140502}
  (\bibinfo{year}{2010}).

\bibitem[{\citenamefont{Schuster et~al.}(2010)\citenamefont{Schuster, Sears,
  Ginossar, {DiCarlo}, Frunzio, Morton, Wu, Briggs, Awschalom, and
  Schoelkopf}}]{schuster_high_2010}
\bibinfo{author}{\bibfnamefont{D.~I.} \bibnamefont{Schuster}},
  \bibinfo{author}{\bibfnamefont{A.~P.} \bibnamefont{Sears}},
  \bibinfo{author}{\bibfnamefont{E.}~\bibnamefont{Ginossar}},
  \bibinfo{author}{\bibfnamefont{L.}~\bibnamefont{{DiCarlo}}},
  \bibinfo{author}{\bibfnamefont{L.}~\bibnamefont{Frunzio}},
  \bibinfo{author}{\bibfnamefont{J.~J.~L.} \bibnamefont{Morton}},
  \bibinfo{author}{\bibfnamefont{H.}~\bibnamefont{Wu}},
  \bibinfo{author}{\bibfnamefont{G.~A.~D.} \bibnamefont{Briggs}},
  \bibinfo{author}{\bibfnamefont{D.~D.} \bibnamefont{Awschalom}},
  \bibnamefont{and} \bibinfo{author}{\bibfnamefont{R.~J.}
  \bibnamefont{Schoelkopf}}, \bibinfo{journal}{Phys. Rev. Lett.}
  \textbf{\bibinfo{volume}{105}}, \bibinfo{pages}{140501}
  (\bibinfo{year}{2010}).

\bibitem[{\citenamefont{Kitaev}(2007)}]{kitaev2007unpaired}
\bibinfo{author}{\bibfnamefont{A.~Y.} \bibnamefont{Kitaev}},
  \bibinfo{journal}{Physics-Uspekhi} \textbf{\bibinfo{volume}{44}},
  \bibinfo{pages}{131} (\bibinfo{year}{2007}).

\bibitem[{\citenamefont{Lutchyn et~al.}(2010)\citenamefont{Lutchyn, Sau, and
  Das~Sarma}}]{lutchyn2010majorana}
\bibinfo{author}{\bibfnamefont{R.~M.} \bibnamefont{Lutchyn}},
  \bibinfo{author}{\bibfnamefont{J.~D.} \bibnamefont{Sau}}, \bibnamefont{and}
  \bibinfo{author}{\bibfnamefont{S.}~\bibnamefont{Das~Sarma}},
  \bibinfo{journal}{Phys. Rev. Lett.} \textbf{\bibinfo{volume}{105}},
  \bibinfo{pages}{077001} (\bibinfo{year}{2010}).

\bibitem[{\citenamefont{Oreg et~al.}(2010)\citenamefont{Oreg, Refael, and von
  Oppen}}]{oreg2010helical}
\bibinfo{author}{\bibfnamefont{Y.}~\bibnamefont{Oreg}},
  \bibinfo{author}{\bibfnamefont{G.}~\bibnamefont{Refael}}, \bibnamefont{and}
  \bibinfo{author}{\bibfnamefont{F.}~\bibnamefont{von Oppen}},
  \bibinfo{journal}{Phys. Rev. Lett.} \textbf{\bibinfo{volume}{105}},
  \bibinfo{pages}{177002} (\bibinfo{year}{2010}).

\bibitem[{\citenamefont{Mourik et~al.}(2012)\citenamefont{Mourik, Zuo, Frolov,
  Plissard, Bakkers, and Kouwenhoven}}]{mourik2012signatures}
\bibinfo{author}{\bibfnamefont{V.}~\bibnamefont{Mourik}},
  \bibinfo{author}{\bibfnamefont{K.}~\bibnamefont{Zuo}},
  \bibinfo{author}{\bibfnamefont{S.~M.} \bibnamefont{Frolov}},
  \bibinfo{author}{\bibfnamefont{S.~R.} \bibnamefont{Plissard}},
  \bibinfo{author}{\bibfnamefont{E.~P. A.~M.} \bibnamefont{Bakkers}},
  \bibnamefont{and} \bibinfo{author}{\bibfnamefont{L.~P.}
  \bibnamefont{Kouwenhoven}}, \bibinfo{journal}{Science}
  \textbf{\bibinfo{volume}{336}}, \bibinfo{pages}{1003} (\bibinfo{year}{2012}).

\bibitem[{\citenamefont{Das et~al.}(2012)\citenamefont{Das, Ronen, Most, Oreg,
  Heiblum, and Shtrikman}}]{das2012zero}
\bibinfo{author}{\bibfnamefont{A.}~\bibnamefont{Das}},
  \bibinfo{author}{\bibfnamefont{Y.}~\bibnamefont{Ronen}},
  \bibinfo{author}{\bibfnamefont{Y.}~\bibnamefont{Most}},
  \bibinfo{author}{\bibfnamefont{Y.}~\bibnamefont{Oreg}},
  \bibinfo{author}{\bibfnamefont{M.}~\bibnamefont{Heiblum}}, \bibnamefont{and}
  \bibinfo{author}{\bibfnamefont{H.}~\bibnamefont{Shtrikman}},
  \bibinfo{journal}{Nature Physics} \textbf{\bibinfo{volume}{8}},
  \bibinfo{pages}{887} (\bibinfo{year}{2012}).

\bibitem[{\citenamefont{Read and Green}(2000)}]{read2000paired}
\bibinfo{author}{\bibfnamefont{N.}~\bibnamefont{Read}} \bibnamefont{and}
  \bibinfo{author}{\bibfnamefont{D.}~\bibnamefont{Green}},
  \bibinfo{journal}{Physical Review B} \textbf{\bibinfo{volume}{61}},
  \bibinfo{pages}{10267} (\bibinfo{year}{2000}).

\bibitem[{\citenamefont{Moore and Read}(1991)}]{moore1991nonabelions}
\bibinfo{author}{\bibfnamefont{G.}~\bibnamefont{Moore}} \bibnamefont{and}
  \bibinfo{author}{\bibfnamefont{N.}~\bibnamefont{Read}},
  \bibinfo{journal}{Nuclear Physics B} \textbf{\bibinfo{volume}{360}},
  \bibinfo{pages}{362} (\bibinfo{year}{1991}).

\bibitem[{\citenamefont{Stern and Halperin}(2006)}]{stern2006proposed}
\bibinfo{author}{\bibfnamefont{A.}~\bibnamefont{Stern}} \bibnamefont{and}
  \bibinfo{author}{\bibfnamefont{B.~I.} \bibnamefont{Halperin}},
  \bibinfo{journal}{Phys. Rev. Lett.} \textbf{\bibinfo{volume}{96}},
  \bibinfo{pages}{016802} (\bibinfo{year}{2006}).

\bibitem[{\citenamefont{Ilan et~al.}(2008)\citenamefont{Ilan, Grosfeld, and
  Stern}}]{ilan2008coulomb}
\bibinfo{author}{\bibfnamefont{R.}~\bibnamefont{Ilan}},
  \bibinfo{author}{\bibfnamefont{E.}~\bibnamefont{Grosfeld}}, \bibnamefont{and}
  \bibinfo{author}{\bibfnamefont{A.}~\bibnamefont{Stern}},
  \bibinfo{journal}{Phys. Rev. Lett.} \textbf{\bibinfo{volume}{100}},
  \bibinfo{pages}{086803} (\bibinfo{year}{2008}).

\bibitem[{\citenamefont{Ilan et~al.}(2009)\citenamefont{Ilan, Grosfeld,
  Schoutens, and Stern}}]{ilan2009experimental}
\bibinfo{author}{\bibfnamefont{R.}~\bibnamefont{Ilan}},
  \bibinfo{author}{\bibfnamefont{E.}~\bibnamefont{Grosfeld}},
  \bibinfo{author}{\bibfnamefont{K.}~\bibnamefont{Schoutens}},
  \bibnamefont{and} \bibinfo{author}{\bibfnamefont{A.}~\bibnamefont{Stern}},
  \bibinfo{journal}{Phys. Rev. B} \textbf{\bibinfo{volume}{79}},
  \bibinfo{pages}{245305} (\bibinfo{year}{2009}).

\bibitem[{\citenamefont{Fu}(2010)}]{fu2010electron}
\bibinfo{author}{\bibfnamefont{L.}~\bibnamefont{Fu}}, \bibinfo{journal}{Phys.
  Rev. Lett.} \textbf{\bibinfo{volume}{104}}, \bibinfo{pages}{056402}
  (\bibinfo{year}{2010}).

\bibitem[{\citenamefont{van Heck et~al.}(2011)\citenamefont{van Heck, Hassler,
  Akhmerov, and Beenakker}}]{van2011coulomb}
\bibinfo{author}{\bibfnamefont{B.}~\bibnamefont{van Heck}},
  \bibinfo{author}{\bibfnamefont{F.}~\bibnamefont{Hassler}},
  \bibinfo{author}{\bibfnamefont{A.~R.} \bibnamefont{Akhmerov}},
  \bibnamefont{and} \bibinfo{author}{\bibfnamefont{C.~W.~J.}
  \bibnamefont{Beenakker}}, \bibinfo{journal}{Phys. Rev. B}
  \textbf{\bibinfo{volume}{84}}, \bibinfo{pages}{180502}
  (\bibinfo{year}{2011}).

\bibitem[{\citenamefont{{Golub} and {Grosfeld}}(2012)}]{golub2012charge}
\bibinfo{author}{\bibfnamefont{A.}~\bibnamefont{{Golub}}} \bibnamefont{and}
  \bibinfo{author}{\bibfnamefont{E.}~\bibnamefont{{Grosfeld}}},
  \bibinfo{journal}{\prb} \textbf{\bibinfo{volume}{86}}, \bibinfo{eid}{241105}
  (\bibinfo{year}{2012}), \eprint{1206.0958}.

\bibitem[{\citenamefont{Koch et~al.}(2007)\citenamefont{Koch, Yu, Gambetta,
  Houck, Schuster, Majer, Blais, Devoret, Girvin, and
  Schoelkopf}}]{koch_charge-insensitive_2007}
\bibinfo{author}{\bibfnamefont{J.}~\bibnamefont{Koch}},
  \bibinfo{author}{\bibfnamefont{T.~M.} \bibnamefont{Yu}},
  \bibinfo{author}{\bibfnamefont{J.}~\bibnamefont{Gambetta}},
  \bibinfo{author}{\bibfnamefont{A.~A.} \bibnamefont{Houck}},
  \bibinfo{author}{\bibfnamefont{D.~I.} \bibnamefont{Schuster}},
  \bibinfo{author}{\bibfnamefont{J.}~\bibnamefont{Majer}},
  \bibinfo{author}{\bibfnamefont{A.}~\bibnamefont{Blais}},
  \bibinfo{author}{\bibfnamefont{M.~H.} \bibnamefont{Devoret}},
  \bibinfo{author}{\bibfnamefont{S.~M.} \bibnamefont{Girvin}},
  \bibnamefont{and} \bibinfo{author}{\bibfnamefont{R.~J.}
  \bibnamefont{Schoelkopf}}, \bibinfo{journal}{Phys. Rev. A}
  \textbf{\bibinfo{volume}{76}}, \bibinfo{pages}{042319}
  (\bibinfo{year}{2007}).

\bibitem[{\citenamefont{Kwon et~al.}(2004)\citenamefont{Kwon, Sengupta, and
  Yakovenko}}]{kwon2004fractional}
\bibinfo{author}{\bibfnamefont{H.~J.} \bibnamefont{Kwon}},
  \bibinfo{author}{\bibfnamefont{K.}~\bibnamefont{Sengupta}}, \bibnamefont{and}
  \bibinfo{author}{\bibfnamefont{V.~M.} \bibnamefont{Yakovenko}},
  \bibinfo{journal}{The European Physical Journal B-Condensed Matter and
  Complex Systems} \textbf{\bibinfo{volume}{37}}, \bibinfo{pages}{349}
  (\bibinfo{year}{2004}).

\bibitem[{\citenamefont{Fu and Kane}(2009)}]{fu2009josephson}
\bibinfo{author}{\bibfnamefont{L.}~\bibnamefont{Fu}} \bibnamefont{and}
  \bibinfo{author}{\bibfnamefont{C.~L.} \bibnamefont{Kane}},
  \bibinfo{journal}{Phys. Rev. B} \textbf{\bibinfo{volume}{79}},
  \bibinfo{pages}{161408} (\bibinfo{year}{2009}).

\bibitem[{\citenamefont{Grosfeld and Stern}(2011)}]{grosfeld2011observing}
\bibinfo{author}{\bibfnamefont{E.}~\bibnamefont{Grosfeld}} \bibnamefont{and}
  \bibinfo{author}{\bibfnamefont{A.}~\bibnamefont{Stern}},
  \bibinfo{journal}{PNAS} \textbf{\bibinfo{volume}{108}},
  \bibinfo{pages}{11810} (\bibinfo{year}{2011}).

\bibitem[{\citenamefont{{Mora} and {Le Hur}}(2012)}]{mora2012low}
\bibinfo{author}{\bibfnamefont{C.}~\bibnamefont{{Mora}}} \bibnamefont{and}
  \bibinfo{author}{\bibfnamefont{K.}~\bibnamefont{{Le Hur}}},
  \bibinfo{journal}{ArXiv e-prints}  (\bibinfo{year}{2012}),
  \eprint{1212.0650}.

\bibitem[{\citenamefont{{Dutt} et~al.}(2013)\citenamefont{{Dutt}, {Schmidt},
  {Mora}, and {Le Hur}}}]{dutt2013strongly}
\bibinfo{author}{\bibfnamefont{P.}~\bibnamefont{{Dutt}}},
  \bibinfo{author}{\bibfnamefont{T.~L.} \bibnamefont{{Schmidt}}},
  \bibinfo{author}{\bibfnamefont{C.}~\bibnamefont{{Mora}}}, \bibnamefont{and}
  \bibinfo{author}{\bibfnamefont{K.}~\bibnamefont{{Le Hur}}},
  \bibinfo{journal}{ArXiv e-prints}  (\bibinfo{year}{2013}),
  \eprint{1301.3434}.

\bibitem[{\citenamefont{Nilsson et~al.}(2008)\citenamefont{Nilsson, Akhmerov,
  and Beenakker}}]{nilsson2008splitting}
\bibinfo{author}{\bibfnamefont{J.}~\bibnamefont{Nilsson}},
  \bibinfo{author}{\bibfnamefont{A.~R.} \bibnamefont{Akhmerov}},
  \bibnamefont{and} \bibinfo{author}{\bibfnamefont{C.~W.~J.}
  \bibnamefont{Beenakker}}, \bibinfo{journal}{Phys. Rev. Lett.}
  \textbf{\bibinfo{volume}{101}}, \bibinfo{pages}{120403}
  (\bibinfo{year}{2008}).

\bibitem[{\citenamefont{Tanaka et~al.}(2009)\citenamefont{Tanaka, Yokoyama, and
  Nagaosa}}]{tanaka2009manipulation}
\bibinfo{author}{\bibfnamefont{Y.}~\bibnamefont{Tanaka}},
  \bibinfo{author}{\bibfnamefont{T.}~\bibnamefont{Yokoyama}}, \bibnamefont{and}
  \bibinfo{author}{\bibfnamefont{N.}~\bibnamefont{Nagaosa}},
  \bibinfo{journal}{Phys. Rev. Lett.} \textbf{\bibinfo{volume}{103}},
  \bibinfo{pages}{107002} (\bibinfo{year}{2009}).

\bibitem[{\citenamefont{Jiang et~al.}(2011)\citenamefont{Jiang, Pekker, Alicea,
  Refael, Oreg, and von Oppen}}]{jiang2011unconventional}
\bibinfo{author}{\bibfnamefont{L.}~\bibnamefont{Jiang}},
  \bibinfo{author}{\bibfnamefont{D.}~\bibnamefont{Pekker}},
  \bibinfo{author}{\bibfnamefont{J.}~\bibnamefont{Alicea}},
  \bibinfo{author}{\bibfnamefont{G.}~\bibnamefont{Refael}},
  \bibinfo{author}{\bibfnamefont{Y.}~\bibnamefont{Oreg}}, \bibnamefont{and}
  \bibinfo{author}{\bibfnamefont{F.}~\bibnamefont{von Oppen}},
  \bibinfo{journal}{Phys. Rev. Lett.} \textbf{\bibinfo{volume}{107}},
  \bibinfo{pages}{236401} (\bibinfo{year}{2011}).

\bibitem[{\citenamefont{Law and Lee}(2011)}]{law2011robustness}
\bibinfo{author}{\bibfnamefont{K.~T.} \bibnamefont{Law}} \bibnamefont{and}
  \bibinfo{author}{\bibfnamefont{P.~A.} \bibnamefont{Lee}},
  \bibinfo{journal}{Phys. Rev. B} \textbf{\bibinfo{volume}{84}},
  \bibinfo{pages}{081304} (\bibinfo{year}{2011}).

\bibitem[{\citenamefont{Ioselevich and
  Feigel’man}(2011)}]{ioselevich2011anomalous}
\bibinfo{author}{\bibfnamefont{P.~A.} \bibnamefont{Ioselevich}}
  \bibnamefont{and} \bibinfo{author}{\bibfnamefont{M.~V.}
  \bibnamefont{Feigel’man}}, \bibinfo{journal}{Phys. Rev. Lett.}
  \textbf{\bibinfo{volume}{106}}, \bibinfo{pages}{077003}
  (\bibinfo{year}{2011}).

\bibitem[{\citenamefont{{Rokhinson} et~al.}(2012)\citenamefont{{Rokhinson},
  {Liu}, and {Furdyna}}}]{rokhinson2012fractional}
\bibinfo{author}{\bibfnamefont{L.~P.} \bibnamefont{{Rokhinson}}},
  \bibinfo{author}{\bibfnamefont{X.}~\bibnamefont{{Liu}}}, \bibnamefont{and}
  \bibinfo{author}{\bibfnamefont{J.~K.} \bibnamefont{{Furdyna}}},
  \bibinfo{journal}{Nature Physics} \textbf{\bibinfo{volume}{8}},
  \bibinfo{pages}{795} (\bibinfo{year}{2012}), \eprint{1204.4212}.

\bibitem[{\citenamefont{Hassler et~al.}(2011)\citenamefont{Hassler, Akhmerov,
  and Beenakker}}]{hassler2011top}
\bibinfo{author}{\bibfnamefont{F.}~\bibnamefont{Hassler}},
  \bibinfo{author}{\bibfnamefont{A.~R.} \bibnamefont{Akhmerov}},
  \bibnamefont{and} \bibinfo{author}{\bibfnamefont{C.~W.~J.}
  \bibnamefont{Beenakker}}, \bibinfo{journal}{New Journal of Physics}
  \textbf{\bibinfo{volume}{13}}, \bibinfo{pages}{095004}
  (\bibinfo{year}{2011}).

\bibitem[{\citenamefont{Bonderson and
  Lutchyn}(2011)}]{bonderson2011topological}
\bibinfo{author}{\bibfnamefont{P.}~\bibnamefont{Bonderson}} \bibnamefont{and}
  \bibinfo{author}{\bibfnamefont{R.~M.} \bibnamefont{Lutchyn}},
  \bibinfo{journal}{Phys. Rev. Lett.} \textbf{\bibinfo{volume}{106}},
  \bibinfo{pages}{130505} (\bibinfo{year}{2011}).

\bibitem[{\citenamefont{{Pekker} et~al.}(2013)\citenamefont{{Pekker}, {Hou},
  {Manucharyan}, and {Demler}}}]{pekker2013proposal}
\bibinfo{author}{\bibfnamefont{D.}~\bibnamefont{{Pekker}}},
  \bibinfo{author}{\bibfnamefont{C.-Y.} \bibnamefont{{Hou}}},
  \bibinfo{author}{\bibfnamefont{V.}~\bibnamefont{{Manucharyan}}},
  \bibnamefont{and} \bibinfo{author}{\bibfnamefont{E.}~\bibnamefont{{Demler}}},
  \bibinfo{journal}{ArXiv e-prints}  (\bibinfo{year}{2013}),
  \eprint{1301.3161}.

\bibitem[{\citenamefont{M\"{u}ller et~al.}(2013)\citenamefont{M\"{u}ller,
  Bourassa, and Blais}}]{blais_majorana_2013}
\bibinfo{author}{\bibfnamefont{C.}~\bibnamefont{M\"{u}ller}},
  \bibinfo{author}{\bibfnamefont{J.}~\bibnamefont{Bourassa}}, \bibnamefont{and}
  \bibinfo{author}{\bibfnamefont{A.}~\bibnamefont{Blais}},
  \bibinfo{journal}{arXiv:1306.1539}  (\bibinfo{year}{2013}).

\bibitem[{\citenamefont{Cottet et~al.}(2013)\citenamefont{Cottet, Kontos, and
  Dou\c{c}ot}}]{cottet_majorana_squeezing_2013}
\bibinfo{author}{\bibfnamefont{A.}~\bibnamefont{Cottet}},
  \bibinfo{author}{\bibfnamefont{T.}~\bibnamefont{Kontos}}, \bibnamefont{and}
  \bibinfo{author}{\bibfnamefont{B.}~\bibnamefont{Dou\c{c}ot}},
  \bibinfo{journal}{Phys. Rev B} \textbf{\bibinfo{volume}{88}},
  \bibinfo{pages}{195415} (\bibinfo{year}{2013}).

\bibitem[{\citenamefont{{Schmidt} et~al.}(2013)\citenamefont{{Schmidt},
  {Nunnenkamp}, and {Bruder}}}]{schmidt2013majorana}
\bibinfo{author}{\bibfnamefont{T.~L.} \bibnamefont{{Schmidt}}},
  \bibinfo{author}{\bibfnamefont{A.}~\bibnamefont{{Nunnenkamp}}},
  \bibnamefont{and} \bibinfo{author}{\bibfnamefont{C.}~\bibnamefont{{Bruder}}},
  \bibinfo{journal}{Phys. Rev. Lett.} \textbf{\bibinfo{volume}{110}},
  \bibinfo{eid}{107006} (\bibinfo{year}{2013}), \eprint{1211.2201}.

\bibitem[{\citenamefont{{Zazunov} et~al.}(2011)\citenamefont{{Zazunov},
  {Yeyati}, and {Egger}}}]{zazunov2011coulomb}
\bibinfo{author}{\bibfnamefont{A.}~\bibnamefont{{Zazunov}}},
  \bibinfo{author}{\bibfnamefont{A.~L.} \bibnamefont{{Yeyati}}},
  \bibnamefont{and} \bibinfo{author}{\bibfnamefont{R.}~\bibnamefont{{Egger}}},
  \bibinfo{journal}{\prb} \textbf{\bibinfo{volume}{84}}, \bibinfo{eid}{165440}
  (\bibinfo{year}{2011}), \eprint{1108.4308}.

\bibitem[{Note1()}]{Note1}
\bibinfo{note}{At this point it can be shown that $\Psi
  _k(n_g=3/4;\varphi )=e^{i\varphi }\Psi _k^*(n_g=1/4;\varphi )$ which leads to
  a partial cancellation of terms in Eq.~\ref {eq:dipoleatdeg} for the
  transition corresponding to $(-)$ and the remaining terms are
  anti-symmetric.}

\bibitem[{\citenamefont{Gambetta et~al.}(2011)\citenamefont{Gambetta, Houck,
  and Blais}}]{gambetta2011tunable_coupling}
\bibinfo{author}{\bibfnamefont{J.~M.} \bibnamefont{Gambetta}},
  \bibinfo{author}{\bibfnamefont{A.~A.} \bibnamefont{Houck}}, \bibnamefont{and}
  \bibinfo{author}{\bibfnamefont{A.}~\bibnamefont{Blais}},
  \bibinfo{journal}{Physical Review Letter} \textbf{\bibinfo{volume}{106}},
  \bibinfo{pages}{030502} (\bibinfo{year}{2011}).

\end{thebibliography}

\newpage

\setcounter{figure}{0}
\setcounter{equation}{0}
\makeatletter
\renewcommand{\thefigure}{S\@arabic\c@figure}
\renewcommand{\theequation}{S\@arabic\c@equation}
\makeatother

\section*{Supplementary Material}

The mesoscopic nature of the Majorana-CPB prescribes certain intricacies in the diagonalization procedure which we now elaborate on. It is a common practice to compose non-local Dirac fermion zero modes $\Gamma_1=\frac{1}{\sqrt{2}}(\gamma_1-i\gamma_2)$, $\Gamma_2=\frac{1}{\sqrt{2}}(\gamma_3-i\gamma_4)$ and similarly $\Gamma_1^\dagger$, $\Gamma_2^\dagger$ through hermitian conjugation. This set of operators satisfies the canonical anti-commutation relations for fermions, $\{\Gamma_i,\Gamma_j^\dagger\}=\delta_{ij}$. In addition we introduce a number operator for each superconductor which we denote by $\hat{n}_i$ ($i=1,2$), which counts the number of Cooper pairs in units of 1, and which is conjugate to the superconducting phase $\varphi_i$, $[\hat{n}_i,\varphi_i]=-i$. Importantly, in topological superconductors $\hat{n}_i$ can assume both integer and half-integer eigenvalues. The latter are associated with the presence of an unpaired electron, counted as half a Cooper pair.

We initialize the system by considering, in the absence of tunneling across the junction, a definite parity of the electron number within each superconductor. The occupation of the Dirac zero mode $N_i=\Gamma_i^\dagger\Gamma_i$ is thus set by the parity of the electron number, $N_i=2 n_i(\rm{mod}\, 2)$. Let us denote the initial state of the two Dirac zero modes within one doublet by $| N_1,N_2\rangle$. For simplicity we also assume that initially $n_1=n_2$ (different choices lead to slightly different formulations but the physical results remain the same up to an overall shift of the gate charge).

Next we turn on the couplings $E_M$ and $E_J$ associated with single-electron tunneling and Cooper-pair tunneling respectively. We construct basis states for the even and odd parity sectors and relative number of Cooper pairs $\hat{n}=\frac{1}{2}(\hat{n}_1-\hat{n}_2)$
\begin{eqnarray}
	&& \left\{e^{i\varphi n}|N_1,N_2\rangle,\,n\in\mathbb{Z}\right\}, \label{eq:even} \\
	&& \left\{e^{i\varphi n}|\bar{N}_1,\bar{N}_2\rangle,\,n\in\mathbb{Z}+\frac{1}{2}\right\}, \label{eq:odd}
\end{eqnarray}
where $\bar{N}_i=1-N_i$ and $\varphi=\varphi_1-\varphi_2$ is the relative phase satisfying $[\hat{n},\varphi]=-i$. The first (second) set of wave functions is periodic (anti-periodic) under a change of $\varphi$ by $2\pi$. The Cooper-pair tunneling operator modifies the eigenvalue of $\hat{n}$ by $\pm 1$, hence only couples states internally within the two subspaces (\ref{eq:even}) and (\ref{eq:odd}). In contrast, the single-electron tunneling operator intermixes the two subspaces. To model the latter we introduce electron raising and lowering operators $e^{\pm i \varphi_i/2}$ satifying $[n_i,e^{\pm i \varphi_i/2}]=\pm \frac{1}{2}e^{\pm i \varphi_i/2}$. These operators are double-valued as they change sign when $\varphi_i\to\varphi_i+2\pi$ and need to be matched with the Majorana zero mode operator which also changes sign, leading to admissible operators (the spectrum is necessarily $2\pi$ periodic although some parity constrained states may exhibit $4\pi$ periodicity \cite{kitaev2007unpaired}). The single electron tunneling term is thus written in terms of the relative phase and the Majorana zero modes as in Eq.~(\ref{eq:maj-tunneling}), with
$
	\nonumber 2i\gamma_2\gamma_3=\Gamma_1^\dagger \Gamma_2+\Gamma_2^\dagger \Gamma_1+\Gamma_1^\dagger\Gamma_2^\dagger+\Gamma_2\Gamma_1
$
fully shuffling the state of the zero mode occupations: $|N_1,N_2\rangle\leftrightarrow |\bar{N}_1,\bar{N}_2\rangle$. In addition, the phase dependent part of the electron tunneling operator ensures that the eigenvalue of $\hat{n}$ changes by $\pm 1/2$. Hence states of the form (\ref{eq:even}) can only couple to states of the form (\ref{eq:odd}).

By projecting on the orthogonal states $|N_1,N_2\rangle$ and $|\bar{N}_1,\bar{N}_2\rangle$ the Hamiltonian acquires a matrix structure, see Eq. \ref{eq:ham}, or in the alternative basis discussed in the main text, Eq. \ref{eq:ham-transformed}.

\begin{figure}[h]
\centering
\includegraphics[scale=0.35]{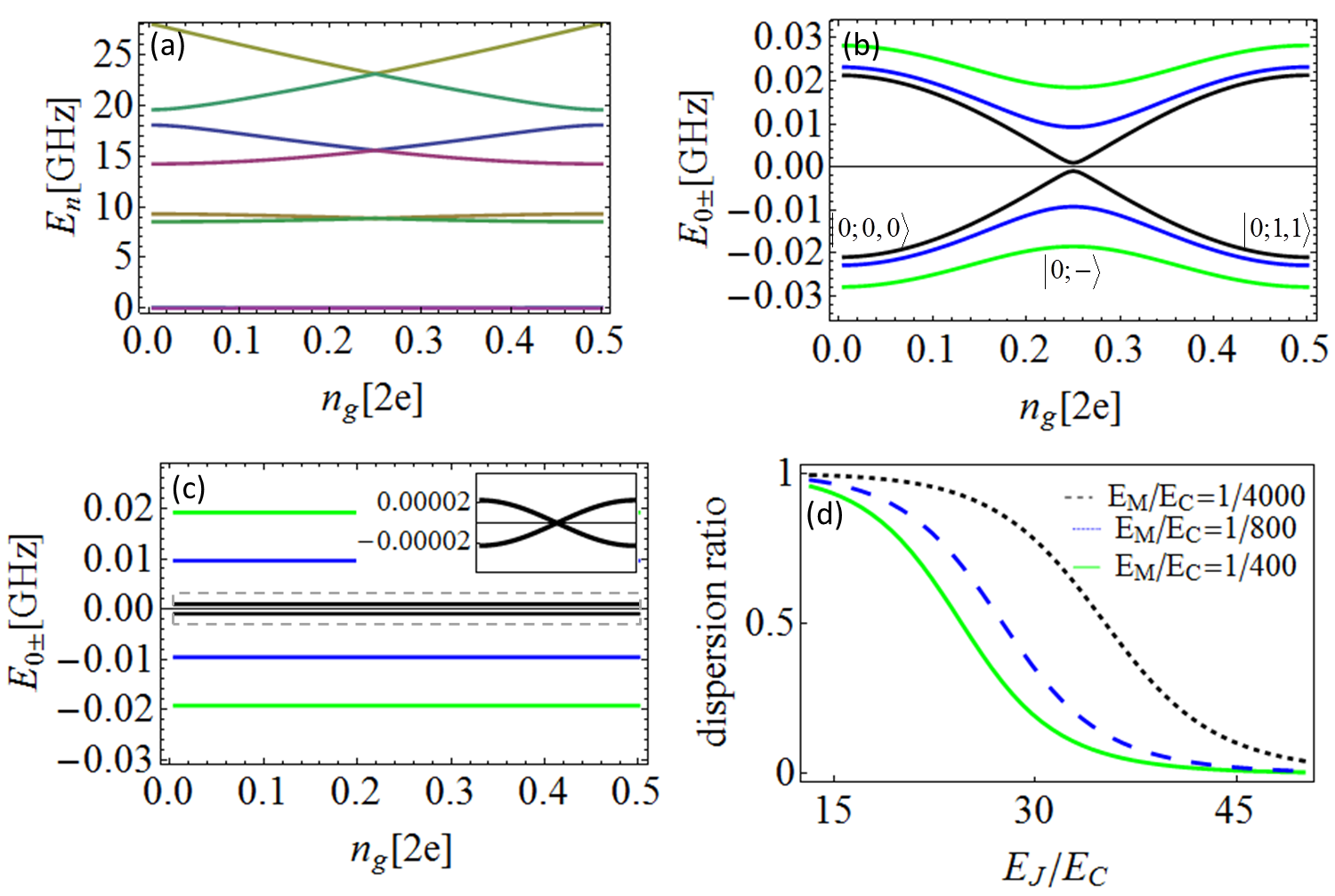}
\caption{(Color online) The different effects of the Majorana interaction on the spectrum. (a) Spectrum of the Majorana-Cooper-Pair-Box spectrum with $E_J/h=10\,\textrm{GHz},E_C/h=0.4\,\textrm{GHz}$ for $E_J/E_C=25$ and $E_M/E_C=1/400$ as a function of the charge offset $n_g$. An avoided crossing develops (b) between the states of different fermion parity and the periodicity of the spectrum is halved, to repeat when $n_g \rightarrow n_g+1/2$. The ground state pair is shown for $E_M/E_C=1/1400$ (black), $E_M/E_C=1/140$ (blue) and $E_M/E_C=1/70$ (green) for $E_J/E_C$=7.14 (with $E_C/h=1.4\,\textrm{GHz}$). (c) For $E_J/E_C=25$ the parity states for $E_M=0$ are crossing but exponentially close in energy with a dispersion of $\epsilon_{0+}(E_M=0)/h=2\times 10^{-5}\,\textrm{GHz}$ (see inset graph of the gray dashed area, with same x-axis). However, including the Majorana coupling $E_M>0$ effectively removes the degeneracy and determines the energy splitting in the whole range of $n_g$. (d) In the transmon regime the residual dispersion is further suppressed by the Majorana interaction $H_M$, as the plot of dispersion ratio $\epsilon_{1+}(E_M)/\epsilon_{1+}(0)$  shows.}
\label{fig:spectrum}
\end{figure}

\end{document}